\documentclass{nature}[a4paper]
\usepackage{bm}
\usepackage{pifont}
\usepackage{marvosym}
\usepackage[colorlinks,linkcolor = blue,citecolor = blue,anchorcolor = blue]{hyperref}
\usepackage{multirow}
\usepackage{multicol}
\usepackage{geometry}
\usepackage{authblk}
\usepackage{amsmath}
\usepackage{graphicx}
\usepackage{xcolor}
\usepackage[numbers,compress]{natbib}

\makeatletter
\renewenvironment{thebibliography}[1]{%
  \section*{\refname}%
  \small
  \list{[\@biblabel{\@arabic\c@enumiv}]}%
       {\settowidth\labelwidth{[\@biblabel{#1}]}%
        \leftmargin\labelwidth
        \advance\leftmargin\labelsep
        \usecounter{enumiv}%
        \let\p@enumiv\@empty
        \renewcommand\theenumiv{\@arabic\c@enumiv}}%
  \sloppy
  \clubpenalty4000
  \widowpenalty4000%
  \sfcode`\.\@m}
{\def\@noitemerr
  {\@latex@warning{Empty `thebibliography' environment}}%
  \endlist}
\renewcommand{\@biblabel}[1]{[#1]}
\makeatother

\makeatletter
\let\saved@includegraphics\includegraphics
\AtBeginDocument{\let\includegraphics\saved@includegraphics}
\renewenvironment{figure}{\@float{figure}}{\end@float}
\makeatother
\usepackage[labelfont={footnotesize}]{caption}
\usepackage{cite}

\makeatletter
\let\oldcite\cite
\renewcommand{\cite}[1]{\mbox{\oldcite{#1}}}
\makeatother

\title{\large Three-dimensional quantum anomalous Hall effect in Weyl semimetals 
\footnotetext{\scriptsize ${}^\ast$ Corresponding author. jianghuaphy@fudan.edu.cn (H.Jiang)}
\footnotetext{\scriptsize ${}^1$  These authors contributed equally to this work: Zhi-Qiang Zhang and Yu-Hang Li.}}

\author[a,b,1]{\small Zhi-Qiang Zhang}
\author[c,d,1]{Yu-Hang Li}
\author[e]{Ming Lu}
\author[b]{Hongfang Liu}
\author[a,f]{Hailong Li}
\author[a,$\ast$]{Hua Jiang}
\author[a,f,g]{X. C. Xie \vspace{-8pt}}

\affil[a]{\scriptsize Interdisciplinary Center for Theoretical Physics and Information Sciences (ICTPIS), Fudan University, Shanghai 200433, China \vspace{-4pt}}
\affil[b]{\scriptsize School of Physical Science and Technology, Soochow University, Suzhou 215006, China \vspace{-4pt}}
\affil[c]{\scriptsize School of Physics, Nankai University, Tianjin 300071, China\vspace{-4pt}}
\affil[d]{\scriptsize State Key Laboratory of the Surface Physics, Fudan University, Shanghai 200433, China\vspace{-4pt}}
\affil[e]{\scriptsize Beijing Academy of Quantum Information Sciences, Beijing 100193, China\vspace{-4pt}}
\affil[f]{\scriptsize International Center for Quantum Materials, School of Physics, Peking University, Beijing 100871, China\vspace{-4pt}}
\affil[g]{\scriptsize Hefei National Laboratory, Hefei 230088, China\vspace{-4pt}}

\graphicspath{{figures/}}

\date{}                   
\begin{document}
	
	\maketitle
	
	\vspace{10pt}
	
\footnotesize
\begin{multicols}{2}
The discovery of the quantum Hall effect in the presence of a relatively strong magnetic field has profoundly inspired the study of topological phase of matter \cite{Hasantopological,qi2011topological,chang2023colloquium}, which not only deepens our understanding of condensed materials beyond the scope of symmetry breaking but also holds significant promise in device application with low or even vanishing energy dissipation. In principle, since the role of magnetic field can be completely replaced by magnetic ordering, quantum Hall effect and its anomalous counterpart, termed quantum anomalous Hall effect, typically appear as complementary pair. These two prominent states can be understood within the same framework of the Thouless-Kohmoto-Nightingale-Nijs (TKNN) theory \cite{thouless1982quantized}, which connects them and their essential properties, such as the quantized Hall resistance $R_{\text{h}}=h/e^2$, with intuitive band topology characterized by the first Chern number $\mathcal{C}$ through the bulk-boundary correspondence \cite{hatsugai1993chern}. While the quantum Hall effect was recently extended to three-dimensional case \cite{Wang20173D}, leading to the observation of the three-dimensional quantum Hall effect with strongly enhanced tunability \cite{Zhang2017quantum,tang2019three}, its anomalous counterpart without external magnetic field in three dimensions still remains a gap in the Hall effect family. {Moreover, although prior studies show that QAHE survives when stacking Chern insulator thin films to the 3D limit when the thickness exceeds the vertical mean free path \cite{Kim2018,Jin2018,Zhao20233D}, the QAHE therein is confined only to the same plane perpendicular to the stacking direction, maintaining 2D physics.}

\vspace{-16pt}

In this work, we demonstrate the realization of the three-dimensional (3D) QAHE in a Weyl semimetal (WSM) \cite{wan2011topological,liu2014discovery,xu2015discovery} and thus fill this outstanding gap. We first confirm this 3D QAHE through the quantized Chern number $\mathcal{C}=1$, then establish its {3D} bulk-boundary correspondence, and finally reaffirm it via the distinctive transport properties. Remarkably, we find that the 3D QAHE hosts two chiral {\textit{surface}} states along one of three spatial directions while a pair of chiral {\textit{hinge}} states along another direction, and the location  of the {\textit{hinge}} states depends sensitively on the Fermi energy. These two types of boundary states are further connected through another perpendicularly propagating chiral surface states, whose chirality is also Fermi energy dependent. Consequently, {the system simultaneously exhibits QAHE along all three spatial directions. Moreover,} depending on the transport direction, its Hall resistance can quantize to $0$, $h/e^2$, or $\pm h/e^2$ when the Fermi energy is tuned across the charge neutral point. {The 3D bulk-boundary correspondence and the coexistence of quantum Hall effect along different direction differentiate our system from stacking Chern insulator thin films, where the QAHE is confined to only one particular plane}. This 3D QAHE not only fills the gap in the Hall effect family but also holds significant potential in device applications.

   \vspace{-16pt}

\begin{figure}[!h]
  	\centering
    	\includegraphics[width=0.9\textwidth]{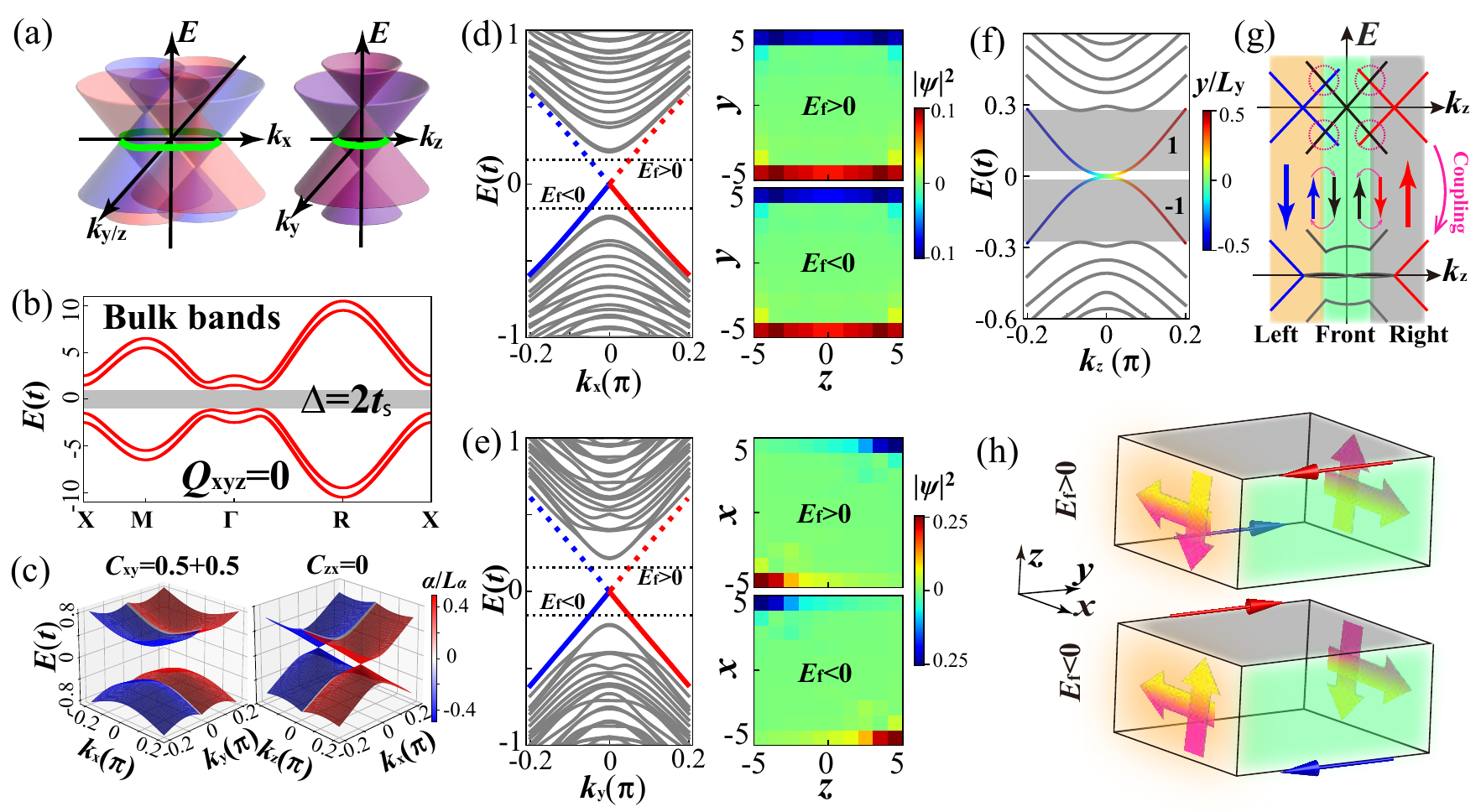}
    	\caption{\scriptsize Construction of 3D QAHE and its bulk-boundary correspondence.
		(a) Schematics for the band dispersions in the absence of spin-orbit coupling {($t_{\text{s}}=0$)}. 
		(b) {Calculated bulk band spectrum of the system along highly symmetric points in the 1st BZ with Rashba spin-orbit coupling ($t_{\text{s}}=0.5t$)}.
		(c) Two-dimensional surface band spectra of a slab and corresponding wavefunction distributions on $k_{\text{x}}$-$k_{\text{y}}$ (left), $k_{\text{z}}$-$k_{\text{x}}$ (right) planes, respectively. {The surface band spectra on the $k_{\text{y}}-k_{\text{z}}$ plane are presented in Supplementary Note 2A}. Here, the thickness of the slab is $L=10$.
		(d,e) Band spectra on a one-dimensional nanowire (left panels) and corresponding wave function distributions. The red and blue colors here represent the propagating direction identical to the group velocity. The system size is $L_{\text{y/x}}=L_{\text{z}}=10$.
		(f) One-dimensional band structure on a nanowire extending infinitely along $z$ direction and the average position $\langle y/L_{\text{y}}\rangle$. The system size is $L_{\text{x}}=L_{\text{y}}=8$. 
        		(g) Mechanism of the emergence of chiral surface states along $z$ direction. 
        		(h) Edge picture of the 3D QAHE at $E_{\text{f}}>0$ (top) and $E_{\text{f}}<0$ (bottom). 
				} 
    	\label{fig1}
\end{figure}

We begin with the following Hamiltonian defined on a cubic lattice
\begin{align}\label{model_Ham}
H=\mathcal{H}_{\text{w}}\sigma_{\text{z}}+\sum_n[it_{\text{s}}(\sigma_{\text{x}}c_n^{\dagger}c_{n+\delta_x}+\sigma_{\text{y}}c_n^{\dagger}c_{n+\delta_y})+\textrm{h.c.}],
\end{align}
where $\mathcal{H}_{\text{w}}=\sum_n\mathcal{T}_0c^\dagger_nc_{n}+\sum_{\alpha= {\text{x,y,z}}}\mathcal{T}_{\alpha}c^\dagger_nc_{n+\delta_\alpha}+{\textrm{h.c.}}$ with $\mathcal{T}_0=-4m\tau_{\text{z}}-\lambda$, $\mathcal{T}_{\text{x}}=m_{\text{x}}\tau_{\text{z}}$, $\mathcal{T}_{\text{y}}=m\tau_{\text{z}}-iA\tau_{\text{y}}$ and $\mathcal{T}_{\text{z}}=m\tau_{\text{z}}-iA\tau_{\text{x}}$ is a typical Hamiltonian for a WSM \cite{liu2014discovery,xu2015discovery}. Here, $\sigma_{\text{x,y,z}}$ ($\tau_{\text{x,y,z}}$) are Pauli matrices acting on spin (pseudospin) spaces, $c_n^\dagger$ ($c_n$) is the electron creation (annihilation) operator at site $n$, $\delta_{\alpha={\text{x,y,z}}}$ is the lattice constant along $\alpha$ direction, which is set to be the length unit, and h.c. refers to the Hermitian conjugate. As shown in Fig.~\ref{fig1}a, the first term in Eq.~\eqref{model_Ham} describes a WSM harboring two pairs of Weyl nodes located at ${\bf{k}}_{\text{w}}^{\pm}=\begin{pmatrix}\pm\pi/2,&0,&0\end{pmatrix}$ and $E_{\pm}=\pm\lambda$. In this case, the Fermi arc band at {$E=0$} forms a closed manifold as indicated by the green lines in Fig.~\ref{fig1}a. The second term is a conventional Rashba spin-orbit coupling with $t_s$ being its strength. Note that the {$\lambda\sigma_z$ term in} Eq.~\eqref{model_Ham} breaks the time-reversal symmetry ($\mathcal{T}$) explicitly, while {the total model Hamiltonian} preserves the space inversion symmetry ($\mathcal{I}$) that can be defined as $\mathcal{I}=I_2\otimes\sigma_{\text{z}}$. {A detailed analysis of the model Hamiltonian construction is provided in Supplementary Note 2A}. The system parameters are hereafter fixed as $m=m_{\text{x}}=t$ and $A=\lambda=t_{\text{s}}=0.5t$ with $t$ the energy unit.

\vspace{-16pt}

Imposing periodic boundary conditions (PBCs) along all three spatial directions allows us to calculate the bulk band spectrum in the momentum space (Supplementary Note 2A), which is displayed in Fig.~\ref{fig1}b along a closed loop connecting all highly symmetric points within the first Brillouin zone (1st BZ). It is evident that the bulk bands are largely gapped with a gap size $\Delta\approx 2t_{\text{s}}$ highlighted by the gray shadow, which excludes the contribution to the topology of the system from these bulk bands. Besides, the bulk topology in three dimensions analogous to 3D QAHE, such as a second-order topological insulator, can be revealed through the octupole moment $Q_{\text{xyz}}$, the $Q_{\text{xyz}}$ turns out to be identically zero and thus removes this possibility (Supplementary Note 2E).  
Fig.~\ref{fig1}c shows the two-dimensional band spectrum on a slab geometry with PBCs along two spatial directions (PBCs along $x$ and $y$ directions in the left panel while along $z$ and $x$ directions in the right panel) while open boundary conditions (OBCs) along the third direction (Supplementary Note 2B). In the left panel, the surface band spectrum on the $x$-$y$ plane is obviously gapped, whose gap minima, in coincidence with the gapless Fermi arc bands in the absence of Rashba spin-orbit coupling (Fig.~\ref{fig1}a), are shifted oppositely along $k_{\text{y}}$. Moreover, the average position $\langle z/L_{\text{z}}\rangle$ that features the location of the wavefunction with respect to the slab layers is superimposed in the same figure. It turns out that  the surface bands with negative $k_{\text{y}}$ are overall confined to the bottom surface at $\langle z/L_{\text{z}}\rangle=-0.5$ (blue region) while those with positive $k_{\text{y}}$ are confined to the opposite top surface at $\langle z/L_{\text{z}}\rangle=0.5$ (red region), manifesting a momentum-surface locking guaranteed by the space inversion symmetry. Remarkably, when the Fermi energy $E_{\text{f}}$ lies inside this gap, each surface band possesses a half-quantized Chern number, leading to a jointly quantized Chern number $\mathcal{C}_{\text{xy}}=0.5+0.5$. In contrast, despite that the surface band in the right panel also manifests a momentum-surface locking,  where negative (positive) $k_{\text{z}}$ bands are confined to the left (right) surface at $\langle y/L_{\text{y}}\rangle=-0.5$ ($\langle y/L_{\text{y}}\rangle=0.5$), the band spectrum on $z$-$x$ plane is gapless with a vanishing Chern number $\mathcal{C}_{\text{zx}}=0$. We also ascertain that the non-zero Berry curvature originates completely from the Fermi arc surface bands (Supplementary Note 2D), further excluding the contribution from the bulk. {Furthermore, the two-dimensional spectrum on the $k_{\text{y}}-k_{\text{z}}$ plane turns out to be gapless, whose Dirac cone lies at $k_{\text{y}}=k_{\text{z}}=0$ (Supplementary Note 2A)}. In view of the bulk-boundary correspondence \cite{hatsugai1993chern}, the quantized Chern number along $z$ direction indicates a lower-dimensional boundary excitation if the system is terminated along $x$ or (and) $y$ directions.
\end{multicols}

\begin{multicols}{2}
 To unveil these boundary states, we further consider a one dimensional nanowire with OBC along $y$ and $z$ directions, and calculate its band spectrum as a function of $k_{\text{x}}$. The results plotted in Fig.~\ref{fig1}d confirm the existence of two gapless boundary states with opposite group velocities (red and blue lines), which extend universally along $z$ direction while propagate oppositely. However, in a nanowire with OBC along $x$ and $z$ directions, the band spectrum in Fig.~\ref{fig1}e shows that the two boundary states are mostly confined to two diagonal hinges of the cross section on the $z$-$x$ plane, and their locations switch from one pair of diagonal hinges to the other pair when the Fermi energy varies across $E=0$, showcasing a pair of energy-dependent chiral hinge states.

\vspace{-16pt}

While the boundary states inherent to the quantized Chern number $\mathcal{C}_{\text{xy}}=1$ have been established, there still exists an unresolved puzzle about the connection between these two different types of boundary states, as their chiralities forbid bending or opposite electron propagating on one surface. To ensure a smooth connection between them, additional vertical chiral states are naturally required, which are anticipated to be also Fermi energy dependent. To uncover hidden boundary states in the 3D QAHE, we further consider a nanowire with OBC along $x$ and $y$ directions. Fig.~\ref{fig1}f displays the band spectrum as a function of $k_{\text{z}}$ and also the average position $\langle y/L_{\text{y}}\rangle$ for four low energy bands close to $E=0$. The wavefunctions of the two central surface bands (colorful lines $\pm1$) are mostly confined to left ($\langle y/L_{\text{y}}\rangle=-0.5$ when $k_{\text{z}}<0$) or right ($\langle y/L_{\text{y}}\rangle=0.5$ when $k_{\text{z}}>0$) surfaces possessing opposite group velocities. These two surface states are protected by a large band gap highlighted by gray shadows, which thus confirms the existence of two vertical chiral surface states that can be ascribed to the coupling among neighboring lateral surface bands. 

\vspace{-16pt}

As illustrated in Fig.~\ref{fig1}g, in the absence of coupling, the gapless Dirac cones on the left and right surfaces are shifted oppositely along $k_{\text{z}}$ while those on the front and back surfaces remain at $k_{\text{z}}=0$. Incorporating the coupling through the common hinge between two neighboring surfaces immediately opens band gaps near the cross regions highlighted by the magenta circles, which simultaneously, eliminates two oppositely propagating surface states denoted by the thin arrows. During this process, only two chiral states separate on left and right surfaces survive. When tuning the Fermi energy across $E=0$, the chirality of these two surface states flips due to the reversal of the group velocities. Consequently, this 3D QAHE altogether hosts two chiral surface states along $x$ direction, a pair of Fermi energy dependent chiral hinge states along $y$ direction, and a pair of chiral surface states along $z$ direction that are also Fermi energy dependent. {This 3D bulk-boundary correspondence, which is schematically summarized in Fig.~\ref{fig1}h, is instructive to understand the anisotropic transport properties of the 3D QAHE}.
\end{multicols}

\begin{figure}[!h]
  	\centering
    	\includegraphics[width=0.9\textwidth]{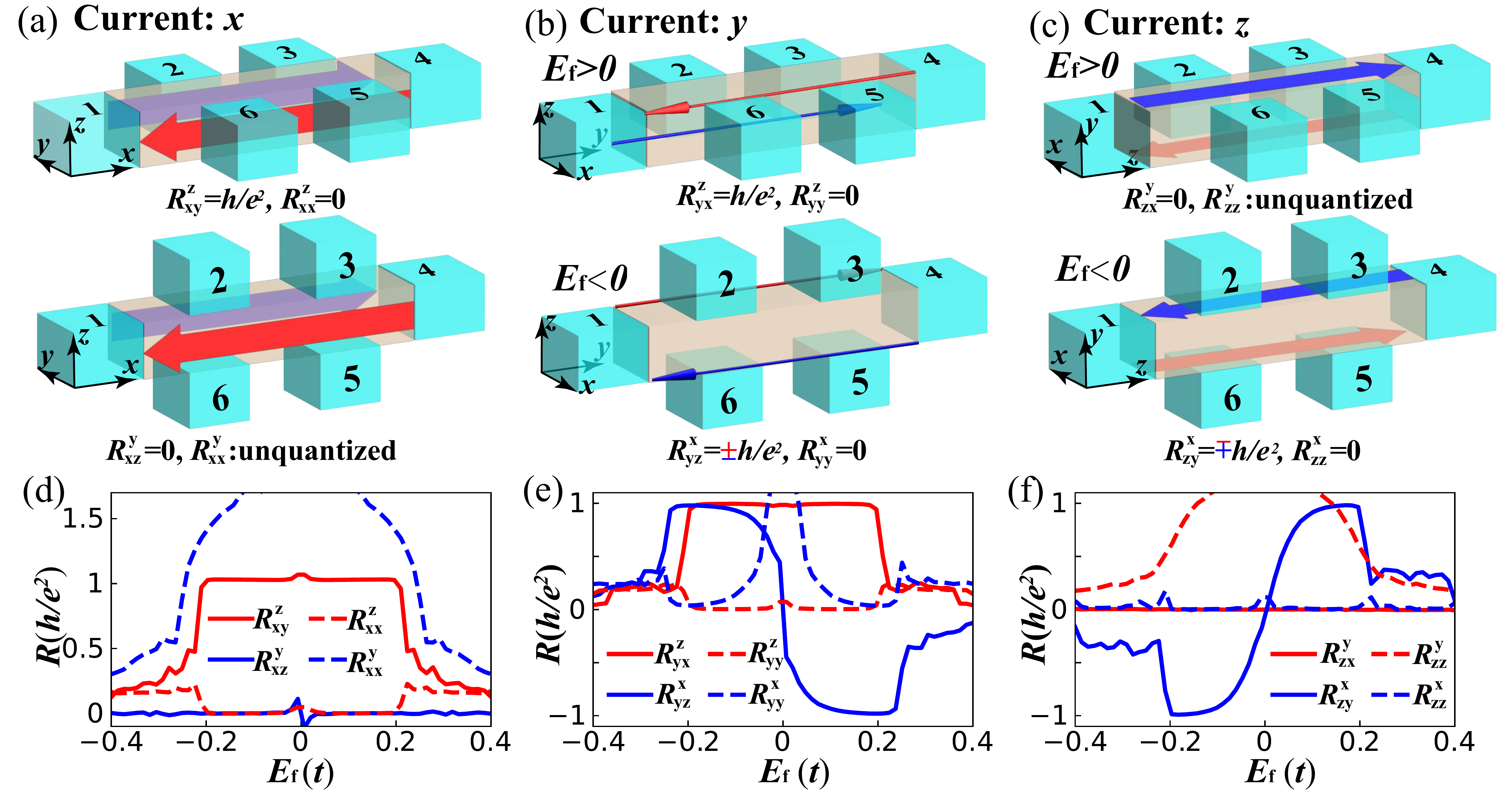}
    	\caption{\scriptsize Transport properties of the 3D QAHE.
		{{(a,b,c)}} Six different {Hall bar} configurations in three dimensions and corresponding edge states (colorful arrows) at different Fermi energy $E_{\text{f}}$. {Note that, for clarity, only the edge states at $E_{\text{f}}>0$ ($E_{\text{f}}<0$) are labeled in top (bottom) panels, respectively, in (b) and (c), while both cases are allowed the same Hall bar configuration}. Hall resistance and longitudinal resistance for each configuration obtained from the edge picture analysis are labeled below.
        {(d,e,f)} Hall resistances (solid lines) and longitudinal resistances (dashed lines) versus the Fermi energy $E_{\text{f}}$ for different Hall bar configurations calculated by using Green's function method.
        Here, the size for all six metallic leads in each Hall bar is $10\times10$, and the distance between neighboring leads is $L=20$. All data are obtained in the presence of a random disorder with strength $W=0.5t$ under 100 times ensemble average.
		} 
    	\label{fig2}
\end{figure}

\begin{multicols}{2}
The crucial fingerprint of the QAHE is the quantized Hall resistance accompanied by a vanishing longitudinal one \cite{chang2023colloquium}, which can be confirmed by the transport measurement in a Hall bar setup with two longitudinal and four transverse leads. To explore the transport properties of the 3D QAHE, we consider a Hall bar structure and calculate its Hall and longitudinal resistances by using the Landauer-B\"uttiker theory \cite{datta1995electronic}. In three dimensions, the current can be applied arbitrarily along all three spatial directions, and in each case, the transverse leads can be attached distinctly along two vertical directions both perpendicular to applied current direction, so there are altogether six geometry allowed Hall bar configurations as depicted in Fig.~\ref{fig2}a-c. For clarity, we denote the Hall and longitudinal resistances in forms of $R^{\alpha}_{\beta\gamma}$ and $R^{\alpha}_{\beta\beta}$, respectively, with $\alpha,\beta,\gamma=x,y,z$, where $\beta$ is the current applied direction, $\gamma$ is the transverse direction while $\alpha$ is the direction perpendicular to both. 

\vspace{-16pt}

 Since the transport in each Hall bar configuration is entirely dominated by involved boundary states, the transport behavior can be obtained from the edge picture. As shown in Fig.~\ref{fig2}a, when the current is applied along $x$ direction, only the chiral {\textit{surface}} states contribute to the transport. If the transverse leads are attached along $y$ direction (top panel), the Hall resistance $R_{\text{xy}}^{\text{z}}$ is anticipated to be quantized to $h/e^2$ while the longitudinal resistance $R_{\text{xx}}^{\text{z}}$ is zero when $E_{\text{f}}$ lies inside the inverted band gap. However, if the transverse leads are attached along $z$ direction (bottom panel), the system should exhibit an insulating behavior with a large $R_{\text{xx}}^{\text{y}}$ and a vanishing $R_{\text{xz}}^{\text{y}}$. When the applied current becomes parallel to $y$ direction, the transport in this case is dominated by the Fermi energy dependent chiral {\textit{hinge}} states as illustrated in Fig.~\ref{fig2}b. If the transverse leads are attached along $x$ direction (top panel), although the hinge states move from one pair of diagonal hinges on the cross section to the other pair when $E_{\text{f}}$ varies across $E=0$, the Hall resistance $R_{\text{yx}}^{\text{z}}$ remains quantized to $h/e^2$ with a vanishing longitudinal resistance $R_{\text{yy}}^{\text{z}}$ since the chirality on $x$-$y$ plane persists. On the contrary, if the transverse leads are attached along $z$ direction (bottom panel), because the chirality of the hinge states on $y$-$z$ plane flips, $R_{\text{yz}}^{\text{x}}$ switches between $\pm h/e^2$ during the manipulation of $E_{\text{f}}$. Finally, when the current is applied along $z$ direction (Fig.~\ref{fig2}c), the transport is dominated by the Fermi energy dependent chiral {\textit{surface}} states { if the transverse leads are attached along $y$-direction (bottom panel)}, whose chirality on $y$-$z$ plane also flips when adjusting the Fermi energy. Therefore, $R_{\text{zy}}^{\text{x}}$ switches between $\pm h/e^2$ with $R_{\text{zz}}^{\text{x}}=0$ (bottom panel) when tuning $E_{\text{f}}$ across $E=0$ whereas $R_{\text{zx}}^{\text{y}}$ is by contrast zero with a large but unquantized $R_{\text{zz}}^{\text{y}}$ (top panel). Our numerical simulations (Fig.~\ref{fig2}d-f) by using the Landauer-B\"uttiker theory (Supplementary Note 3) agree remarkably well with above analysis. Meanwhile, we also confirm that the quantized Hall resistance and the vanishing longitudinal resistance are robust against Anderson disorders, reaffirming the topological nature of this 3D QAHE. Furthermore, we also find highly quantized Chern number state (Supplementary Note 1 and 7).

\vspace{-16pt}

In conclusion, we have proposed a 3D QAHE existing on the surface of a WSM when incorporating a conventional Rashba spin-orbit coupling. This 3D QAHE hosts two chiral surface states along $x$ direction, a pair of Fermi energy dependent chiral hinge states along $y$ direction, and also a pair of Fermi energy dependent chiral surface states along $z$ direction. Those peculiar boundary states inherent to the 3D QAHE in turn results in a highly anisotropic transport behavior along different directions, where the Hall resistance can be zero, $h/e^2$, or $\pm h/e^2$ at different Fermi energies. {Those transport signature makes our proposal fundamentally different from the stacking Chern insulator thin films, which become a 3D materials if the thickness is larger than the vertical mean free path. However, in this scenario, the QAHE is still confined to one particular plane perpendicular to the magnetization, maintaining a quasi-two-dimensional feature}. {Moreover,} the novel transport properties make this 3D QAHE an ideal platform for in-memory computing (Supplementary Note 6). 

\vspace{-16pt}

The realization of this 3D QAHE relies heavily on the incorporation of Rashba spin-orbit coupling into the WSM harboring two pairs of shifted Weyl nodes. Given that many WSMs have been proposed and also confirmed in experiment, our theory could possibly be tested in magnetically doped type-I WSMs, such as Fe, Cr, V doped Cd$_3$As$_2$ or Na$_3$Bi (Supplementary Note 8), which possess two pair of Weyl nodes. {Sizable in-plane spin-orbit coupling aligning with the form in Eq.~(\ref{model_Ham}) can be achieved via the uncompensated internal electric field due to the slight displacement of doped atoms}. Our work extends the QAHE to 3D case, paving the way for the application of topological materials.

\noindent\textbf{Conflict of interest}\vspace{-12pt}

  	The authors declare that they have no conflict of interest.
	
\noindent\textbf{Acknowledgements}\vspace{-12pt}

    We are grateful for the fruitful discussions with Ming Gong, Chui-Zhen Chen and Haiwen Liu. This work is financially supported by the National Basic Research Program of China (Grants Nos. 2024YFA1409003 and 2022YFA1403700), National Natural Science Foundation of China (Grants Nos. 12204044, 12147126 and 12404056). Yu-Hang Li also acknowledge financial support from the Fundamental Research Funds for the Central Universities, the State Key Laboratory of Surface Physics and the Department of Physics at Fudan University.
    
\noindent\textbf{Author contributions}\vspace{-12pt}

    Hua Jiang and X.C. Xie conceived the initial idea of 3D QAHE. Zhi-Qiang Zhang and Yu-Hang Li performed calculations with assistance from Ming Lu. All authors discussed the results. Zhi-Qiang Zhang and Yu-Hang Li wrote the manuscript with contributions from all authors. Hua Jiang and X.C. Xie supervised the project.
    
\noindent\textbf{Appendix A. Supplementary materials}\vspace{-12pt}
        
    Supplementary materials to this short communication can be
found online at


\noindent\textbf{References}
\begin{thebibliography}{15}

\bibitem{Hasantopological}
Hasan MZ, Kane CL, Colloquium:Topological insulators. \emph{Rev Mod Phys} 2010;82:3045-3047. \vspace{-8pt}

\bibitem{qi2011topological}
Qi XL, Zhang SC. Topological insulators and superconductors. \emph{Rev Mod Phys} 2011;83:1057-1110. \vspace{-21pt}

\bibitem{chang2023colloquium}
Chang QY, Li XJ, Zhang YJ, et al. Colloquium: Quantum anomalous Hall effect. \emph{Rev Mod Phys} 2023;95:011002. \vspace{-8pt}

\bibitem{thouless1982quantized}
Thouless DJ, Kohmoto M, Nightingale MP, et al. Quantized Hall conductance in a two-dimensional periodic potential. \emph{Phys Rev Lett} 1982;49:405. \vspace{-8pt}

\bibitem{hatsugai1993chern}
Hatsugai Y. Chern number and edge states in the integer quantum Hall effect. \emph{Phys Rev Lett} 1993;71:3697. \vspace{-8pt}

\bibitem{Wang20173D}
Wang CM, Sun HP, Lu HZ, et al. 3D quantum Hall effect of Fermi arcs in topological semimetals.\emph{Phys Rev Lett} 2017;119:136806. \vspace{-8pt}

\bibitem{Zhang2017quantum}
Zhang C, Zhang Y, Yuan X, et al, Quantum Hall effect based on Weyl orbits in {Cd${}_3$As${}_2$}. \emph{Nature} 2019;565:331–336. \vspace{-8pt}

\bibitem{tang2019three}
Tang F, Ren Y, Wang P, et al. Three-dimensional quantum Hall effect and metal–insulator transition in {ZrTe${}_5$}. \emph{Nature} 2019;569:537–541. \vspace{-8pt}

\bibitem{Kim2018}
Kim SW, Seo K, Uchoa B, Three-dimensional quantum anomalous Hall effect in hyperhoneycomb lattices. \emph{Phys Rev B} 2018;97:201101(R). \vspace{-8pt}

\bibitem{Jin2018}
Jin YJ, Wang R, Xia BW, Three-dimensional quantum anomalous Hall effect in ferromagnetic insulators. \emph{Phys Rev B} 2018;98:081101(R). \vspace{-8pt}

\bibitem{Zhao20233D}
Zhao YF, Zhang R, Sun ZT, et al, 3D Quantum anomalous Hall effect in magnetic topological insulator trilayers of hundred-nanometer thickness. \emph{Adv Mater} 2024;36:2310249. \vspace{-8pt}

\bibitem{wan2011topological}
Wan X, Turner AM, Vishwanath A, et al. Topological semimetal and Fermi-arc surface states. \emph{Phys Rev B} 2011;83:205101. \vspace{-8pt}

\bibitem{liu2014discovery}
Hasan M, Xu SY, Belopolski I, et al, Discovery of Weyl fermion semimetals and topological Fermi arc states. \emph{Annu Rev Condens Matter Phys} 2017;8:289–309. \vspace{-8pt}

\bibitem{xu2015discovery}
Xu SY, Belopolski I, Alidoust N, et al. Discovery of a Weyl fermion semimetal. \emph{Science} 2015;349:613–617. \vspace{-8pt}

\bibitem{datta1995electronic}
Datta S. Electronic transport in mesoscopic systems. \emph{Cambridge University Press}, Cambridge UK, 1995.
\end{thebibliography}

\end{multicols}
    
\end{document}